# TEXT-BASED EMOTION AWARE RECOMMENDER

John Kalung Leung[1], Igor Griva[2] and William G. Kennedy[3]

[1]Computational and Data Sciences Department, Computational Sciences and Informatics, College of Science, George Mason University, 4400 University Drive, Fairfax, Virginia 22030, USA
[2]Department of Mathematical Sciences, MS3F2, Exploratory Hall 4114, George Mason University,4400 University Drive, Fairfax, Virginia 22030, USA
[3]Center for Social Complexity, Computational and Data Sciences Department, College of Science, George Mason University, 4400 University Drive, Fairfax, Virginia 22030, USA

## ABSTRACT

*We apply the concept of users' emotion vectors (UVECs) and movies' emotion vectors (MVECs) as building components of Emotion Aware Recommender System. We built a comparative platform that consists of five recommenders based on content-based and collaborative filtering algorithms. We employed a Tweets Affective Classifier to classify movies' emotion profiles through movie overviews. We construct MVECs from the movie emotion profiles. We track users' movie watching history to formulate UVECs by taking the average of all the MVECs from all the movies a user has watched. With the MVECs, we built an Emotion Aware Recommender as one of the comparative platforms' algorithms. We evaluated the top-N recommendation lists generated by these Recommenders and found the top-N list of Emotion Aware Recommender showed serendipity recommendations.*

## 1. INTRODUCTION

We have illustrated in the paper [1] the benefit of using movie emotion vectors (mvec) and user emotion vectors (uvec) to enhance a Recommender's top-N recommendation making process. The goal of this paper is to make use of mvec and uvec embeddings as emotional components besides making the top-N recommendations, also develop an end-to-end Emotion Aware Recommender (EAR). In the article, [1], the mvec embeddings represent a movie's emotional features derived from the movie overview. We developed a Tweets Affective Classifier (TAC) cable of classifying six primary human emotions, and we added a neutral mood to TAC for affective computing convenience. We use TAC to classify movie overviews to obtain the movie's emotional profile, mvec. A uvec embeddings represent the mean value of all the mvec film moods' embeddings a user has watched. In this paper, we expand the coverage of the mvec embeddings to include other movies' textual metadata, such as genres. We denote the expanded mvec embeddings as item vectors (ivec). In the same token, we named the extended coverage of uvec, wvec.

We demonstrated, in [1], the affective movie recommendation making through an SVD-CF Recommender. In this study, we build a comparative Recommender platform, which makes movie recommendations through Recommender algorithms of Content-based (CB), and Collaborative Filtering (CF). In the case of CB Recommender, we develop a movie genres CB Recommender denotes as Genres Aware Recommender (GAR). We transform mvec embeddings of movie overviews into a multi-label emotion classification in One-Hot Encoded (OHE) embeddings and named the embeddings as ivec. We build an ivec embeddings CB Recommender and denote it as Emotion Aware Recommender (EAR). We then combine the emotion and genres into an expanded ivec for developing a Multi-channel Aware Recommender (MAR). We also construct an Item-based Collaborative Filtering (IBCF) and a User-based Collaborative Filtering (UBCF) Recommenders from scratch. We compare the differences between the five recommender algorithms' comparative performance through the recommendations-making process.

We apply the Cosine Similarity depicted in equation 2 as the primary algorithm in building our Recommender platform. In the case of recommended movies in the top-N recommendations contain similar genres of films that the active user has watched and liked, Cosine Similarity will reveal the closeness in the similarity between the recommended movies and the movies the active user has viewed and liked. Similarly, we can apply Cosine Similarity to find the similarity in the emotion profile of a top-N movies list and the movie's emotion profile of an active user who has watched and loved. Moreover, in UBCF, we apply a rating matrix, $R$, to compute the collaborative filtering for recommending movies to an active user. Each row of UBCF in $R$ represents the rating value of films a user has watched and rated; whereas, each column in $R$ represents a movie of rating scores it received from users who have viewed and assessed. By comparing the Cosine Similarity between the active user and a user in the corresponding rows, effectively, we compare two rows in $R$; thus, we know the closeness of the two users. Once we find the closest similarity score of the active user and a particularuser in $R$, we scan the active user's unwatched movies that match the watched films of the closest similar user. Through collaborative filtering, we make the top-N movie recommendations to the active user. Lastly, we evaluate the performance of Recommenders in the comparative platform by contrasting each top-N recommendation list generated by the five Recommender algorithms. We find the top-N recommendation list made by the Emotion Aware Recommender (EAR) shows intrigue results.

In the advent of the Internet era, large conglomerates, small and medium businesses (SMB), have deployed Recommender Systems to gain a competitive advantage in providing customers a personalized, useful business transaction experience while understanding customers' tastes and decision-making habits. For customers who left feedback regarding their experiences of the goods and services they received, Recommender can mine customers' opinions through sentiment analysis (SA) to better understand the what, why, and how customers' likes and dislikes the goods and services they consumed. Also, if customers have rated the goods and services, Recommender can make use of the rating information along with the sentiment analysis on opinion feedbacks to make a future personalized recommendation of products and services to customers that meet their tastes and expectation. For example, such Recommender is known as Hybrid Recommender System using Collaborative Filtering with Sentiment Analysis (CF-SA) [2]. CF-SA Recommender is also known to outperform the baseline Collaborate Filtering Recommender System in personalized recommendation making [3] [4].

Nevertheless, no Recommender was built with design to explicitly collect human emotions data [5] [6]. Also, no publicly available dataset contains explicit affective features for implementing a Recommender System. The alternative for Recommender researchers is to build an affective aware Recommender by deriving the needed emotional features from some datasets implicitly [7] and [8] [9]. Movies and music datasets are the two most popular datasets with metadata, such as genres and reviews for affective features mining [5].

In the next section, we will level set readers with the related work in the field of affective computing and Emotion Aware Recommenders. Next, we will illustrate the development of the comparative platform for the five Recommender algorithms, Tweets Affective Classifier, and datasets in the methodology section. Next, in the implementation section, we will highlight the five Recommenders with flowcharts. In the evaluation section, we will show the top-N recommendations lists generated by the five Recommender algorithms while contrasting their differences. We also will highlight our observations regarding the limitations and deficiencies of developing the comparative platform. We will document our future work plan in the future work section before closing our report with a conclusion. Following the conclusion section is the reference section and the authors' brief biography.

## 2. RELATED WORK

Emotion Aware Recommender System (EAR) is a field in active research. Illustrated below are samples of a few recent works. Orellana-Rodriguez [10] [11] advocated that instead of detecting the affective polarity features (i.e., positive/negative) of a given short video in YouTube, they detect the paired eight basic human emotions advocated by Plutchik [12] [13] into four opposing pairs of basic moods: joy–sadness, anger–fear, trust–disgust, and anticipation–surprise. Orellana-Rodriguez [10] also leveraged the auto extraction of film metadata's moods context for making emotion-aware movie recommendations. Qian et al. [14] proposed an EARS based on hybrid information fusion using user rating information as explicit data, user social network data as implicit information, and sentiment from user reviews as the source of emotional information. They [14] also claimed the proposed method achieved higher prediction ratings and significantly enhanced the recommendation accuracy. Also, Narducci et al. [15] [16] described a general architecture for building an EARS and demonstrated through a music Recommender with promising results.

Moreover, Mizgajski and Morzy [17] formulated an innovative multi-dimensional model EARS for making recommendations on a large-scale news Recommender. The database consists of over 13 million news pages based on 2.7 million unique user's self-assessed emotional reactions resulted in over 160,000 emotional reactions collected against 85,000 news articles. Katarya and Verma [5] completed a literature review of research publications in the Affective Recommender Systems (ARS) field from 2003 to February 2016. The report offers in-depth views of the evolution of technology and the development of ARS.

The field of human primary Emotion Detection and Recognition (EDR) through artificial intelligence methods is in active research [18] [19] [20] [21] [22] [23] [24] [25]. In the case of image-oriented data, Facial Detection, and Recognition (FDR) is the main thrust in research [26] [27] to study basic human emotions through facial expression. For textual based data with subjective writing, Sentiment Analysis (SA) takes the lead [28] [29] [30] to extract emotions from fine-grained sentiment. The aim is to uncover the affective features from texts or images and classify the emotional features into the categories of moods. Paul Ekman, a renowned psychologist and professor emeritus at the University of California, San Francisco, advocated the six basic human moods classification: happiness, sadness, disgust, fear, surprise, and anger [31] [32]. Ekman later added "contempt" as the seventh primary human emotion to his list [33] [34]. Another renowned psychologist, Robert Plutchik, invented the Wheel of Emotions advocated eight primary emotions: anger, anticipation, joy, trust, fear, surprise, sadness, and disgust [12]. Research at Glasgow University in 2014 amended that couple pairs of primary human emotions such as fear and surprise elicit similar facial muscles response, so are disgust and anger. The study broke the raw human emotions down to four fundamental emotions: happiness, sadness, fear/surprise, and disgust/anger [35] [36]. This paper adopts Paul Ekman's classification of six primary human emotions: happiness, sadness, disgust, fear, surprise, and anger for modeling the ivec embeddings while adding "neutral" as the seventh emotion feature for convenience in affective computing.

FDR on facial expression has a drawback - it fails to classify an image's emotional features with the absence of human face on the image. In the case of using FDR to classify movie poster images, often, the poster may contain a faceless image. Thus, we propose to indirectly classify the affective features of a poster image through textual-based emotion detection and recognition (EDR) using a movie overview rather than facial-based FDR directly on the poster image.

## 3. METHODOLOGY

We propose an innovative method as our contribution to Recommender research, which based on the following sources:

- item's explicit rating information
- item's implicit affective data embeddings
- user's emotion and taste profile embedding

To implement an end-to-end Multi-channel Emotion Aware Recommender System (e2eMcEARS) or McEAR for short. Several researchers have documented that emotions playing an essential role in the human decision-making process [37] [38] [39] [40] [41] [42]. Also, psychologists and researchers in social science know that the state of mind or moods of an individual affects his decision-making processes [43] [44] [45] [46]. We envision that affective embeddings can represent any product or service. In our previous work [1], we illustrated a method to derive an emotion classifier from tweets' affective tags and use the affective model to predict the mood of a movie through the movie overview. We denoted the mood embeddings of the movie as mvec. We also stated that the value of the embedding of a mvec would hold the same value throughout its lifespan. Also, we denote uvec represents the average value of all mvec of the movies a user has watched. The value of uvec will change each time the user watches a movie. We want to expand the coverage of the mvec to other metadata of the movie, such as genres. We denote the expanded mvec as item embeddings (ivec), which holds the mood embeddings of movie overview and genres. Similarly, uvec will expand its embedding as the average value of all ivec of the movies a user has consumed. We denote the expanded uvec embeddings as wvec.

### 3.1. Overview of the Tweets Affective Classifier Model

We developed the Tweets Affective Classifier (TAC), as illustrated in [1], which employed an asymmetric butterfly wing double-decker bidirectional LSTM - CNN Conv1D architecture to detect and recognize emotional features from tweets' text messages. We have preprocessed the seven emotion words embeddings to be used as input to train TAC through the pre-trained GloVe embeddings using the glove.twitter.27B.200d.txt dataset. We have two types of input words embeddings: trainable emotion words embeddings and frozen emotion words embeddings. By frozen the embeddings, we mean the weights in the embeddings are frozen and cannot be modified during TAC's training session. We started with the first half of the butterfly wing by feeding preprocessed TAC input emotion words embeddings to the double-decker bidirectional LSTM neural nets. We fed the frozen emotion words embeddings to the top bidirectional LSTM and fed the trainable emotion words embeddings to the bottom bidirectional LSTM. Next, we concatenated the top and bottom bidirectional LSTM to form the double-decker neural net. We fed the output from the double-decker bidirectional LSTM to seven sets of CNN Conv1D neural nets with the dropout parameter set at 0.5 in each set of Conv1D as regularization to prevent the neural net from overfitting. We then concatenated all the outputs of Conv1Ds to form the overall output of the first half of the butterfly wing neural nets.

The architecture layout of the second half of the butterfly wing neural nets is different from the peer. We started by setting up seven pairs of CNN Conv1D neural nets. With each pair of Conv1D, we fed in parallel the preprocessed TAC's frozen emotion words embeddings as input to a Conv1D and the trainable emotion words embeddings to the other. We set the dropout parameter at 0.5 for all seven pairs of conv1D to prevent overfitting. We concatenated all the outputs of seven pairs Conv1D to become a single output and fed that to a single bidirectional LSTM with the dropout value set at 0.3. We then concatenated the first half of the butterfly wing output with the second half to form the overall output. Next, the output then fed through in series to a MaxPooling1D with the dropout value set at 0.5, followed by a Flatten neural net before going through a Dense neural net and another Dense neural net with sigmoid activation to classify the emotion classification in a probabilistic distribution. When predicting a movie's emotion profile using TAC, the classifier will classify the mood of a movie through the movie overview. TAC output the movie emotion prediction in the form of the probabilistic distribution of seven values, indicating the value in percentage of each class of the seven emotions, or the emotion profile of the movie.

## 3.2. Overview of Comparative Platform for Recommenders

Building the comparative platform for Recommenders from scratch provides a way to study and observe the process of making recommendations under different context situations. We apply the most basic method to build the collection of Recommenders in the comparative platform. Thus, we are not aiming for best practice algorithms to build Recommender with a high performance nor high throughput in mind; but it is easy to modify and adapt to a different information context, and highly functional is most desirable. A Recommender is known to build with a specific domain in mind. As we march down the path of researching Emotion Aware Recommenders, we want the comparative platform that we are developing for the movie-oriented Recommenders can later transfer the learning to other information domains.

We reckon that in the context of movie domain, for example, a Genres Aware Recommender (GAR) may be adequate for making movie recommendations through movie genres, but without some adaptable in processing logic, the movie GAR may not handle well when feeding it with music genres. Of course, movie GAR will fail to make recommendations if we feed other domain data absence of genre information. However, primary human emotions are the same universally in different races and cultures. Once we obtain an emotion profile of a user obtains from a domain, the user's same emotion profile should be transferable to other domains with no required modification. The caveat is that the other domain must contain data that is emotion detectable and recognizable or emotion aware enable.

## 3.3. Datasets

The success of any machine learning project requires large enough domain-specific data for computation. For movie-related affecting computing, no affective labeled dataset is readily available. Thus, we need to build the required dataset by deriving it from the following sources. For movie rating datasets, we obtained these datasets from the GroupLens' MovieLens repository [47]. We scraped The Movie Database (TMDb) [48] for movie overviews and other metadata. MovieLens contains a "links" file that provides cross-reference links between MovieLens' movie id, TMDb's tmdb id, and IMDb's imdb id. We connect MovieLens and TMDb datasets through the "links" file.

Using a brute force method, we scrape the TMDb database for movie metadata, particularly for movie overview or storyline, which contains subjective writings of movie descriptions that we can classify the mood of the text. We can query the TMDb database by tmdb id, a unique movie identifier assigned to a movie. The tmdb id starts from 1 and up. However, in the sequence of tmdb id, gaps may exist between consecutive numbers. Our scraping effort yields 452,102 records after the cleansing of raw data that we scraped from TMDb.

We developed a seven text-based emotion classifier capable of classifying seven basic human emotions in tweets, as illustrated in [1]. We apply the Tweets Affective Classifier (TAC) to classify the moods of movie overviews by running TAC through all the 452,102 overviews that scraped from the TMDb database to create a movie emotion label dataset.

MovieLens datasets come in different sizes. We work with the following MovieLens datasets: the ml-20m dataset, 20 million rating information; the ml-latest-small dataset, about ten thousand rating information of 610 users; ml-latest-full dataset, holds 27 million rating information; and the recently leased ml-25m dataset, with 25 million rating information. The name of the MovieLens dataset coveys the number of ratings, movies, users, and tags contained in the dataset. Table 1 depicts the number of ratings, users, and movies; each of the MovieLens datasets contain. Each of the depicted MovieLens datasets provides a links file to cross-reference between MovieLens and two other movie databases, TMDb and the Internet Movie Database (IMDb ) [49], through movie id, tmdb id, and imdb id. MovieLens maintains a small number of data fields, but users can link it to TMDb and IMDb databases via the links file to access other metadata that MovieLens lacks.

Table 1: MovieLens datasets.

| Datasets | Ratings | Users | Titles |
|----------|---------|-------|--------|
| ml-20m | 20M | 138000 | 27000 |
| ml-25 | 25M | 162000 | 62000 |
| ml-latest-small | 100K | 600 | 9000 |
| ml-latest-full | 27M | 280K | 58000 |

The ml-latest-full dataset is the largest in the MovieLens dataset collection. However, the ml-latest-full dataset will change over time and is not proper for reporting research results. We use the ml-latest-small, and ml-latest-full datasets in proof of concept and prototyping, not research reporting work. The other MovieLens 20M and 25M datasets are stable benchmark datasets which we will use for research reporting.

Although we have scraped 452,102 movie overviews from TMDb when merging with MovieLens, we can only make use of one-eighth of the number of overviews that we have collected. Table 2 shows the number of movie overviews the MovieLens datasets can extract from TMDb after cleaning from raw data.

Table 2: Number of overview in MovieLens extracted from TMDb.

| Datasets | No. of Overviews |
|----------|------------------|
| ml-20m | 26603 |
| ml-25m | 25M |
| ml-latest-small | 9625 |
| ml-latest-full | 56314 |

We merged the MovieLens datasets with the emotion label datasets obtained from TAC. Form our cleansed ml-latest-small training dataset of 9625 rows extracted from the raw 9742 rows, after merging with the emotion label dataset, the applicable data point row is down to 9613. MovieLens datasets are known for preprocessed and cleaned datasets. Nevertheless, when going through the necessary data preparation steps, we still experienced a 1.32% data loss from the original dataset. Depicted below in table 3 is the first few rows of the final cleansed training dataset.

Table 3: First few rows of cleansed training dataset

| Index | tid | mid | iid | mood | neutral | appy | sad | hate | anger | disgust | surprise |
|-------|-----|-----|-----|------|---------|------|-----|------|-------|---------|----------|
| 1 | 2 | 4470 | 94675 | disgust | 0.157 | .086 | 0.156 | 0.075 | 0.085 | 0.266 | 0.175 |
| 2 | 5 | 18 | 113101 | disgust | 0.121 | .060 | 0.098 | 0.128 | 0.133 | 0.244 | 0.216 |
| 3 | 6 | 479 | 107286 | hate | 0.075 | .114 | 0.054 | 0.433 | 0.095 | 0.128 | 0.100 |
| 4 | 11 | 260 | 76759 | neutral | 0.299 | .262 | 0.079 | 0.030 | 0.017 | 0.083 | 0.230 |
| 5 | 12 | 6377 | 266543 | surprise | 0.150 | .080 | 0.055 | 0.083 | 0.103 | 0.153 | 0.376 |

## 4. IMPLEMENTATION

### 4.1. Recommender Platform

We develop a movie Recommender platform for our study to evaluate five Recommender algorithms in movie recommendations making. We employ the following five Recommender algorithms in the Recommender platform.

- an Item-based Collaborative Filtering (IBCF) movie Recommender to compute pairwise items Cosine Similarity, as depicts in equation 2 for identifying the closeness of similar items. The rating matrix, $R$, configures with rows representing movie titles, and columns representing users.
- a User-based Collaborative Filtering (UBCF) movie Recommender to compute pairwise users Cosine Similarity, as depicts in equation 2 for identifying the closeness of similar users. The rating matrix, $R$, configures with rows representing users, and columns representing movie titles.
- A genre-aware Content-based Recommender (GAR) using Cosine Similarity as depicted in equation 2 to compute the pairwise similarity between two movies' genres.
- an emotion aware Content-based Recommender (EAR) using Cosine Similarity as defined in equation 2 to compute the pairwise similarity between two emotion aware movies
- an emotion and genres aware multi-modal Content-based Recommender (MAR) using Cosine Similarity as depicted in equation 2 to compute the pairwise similarity between two items with affective awareness and genres embeddings.

$$Inner(x, y) = \sum_i x_i y_i = <x, y> \qquad (1)$$

$$CosSim(x, y) = \frac{\sum_i x_i y_i}{\sqrt{\sum_i x_i^2} \sqrt{\sum_i y_i^2}} = \frac{<x, y>}{||x|| \, ||y||} \qquad (2)$$

We deployed the MovieLens ml-latest-small dataset as the training set and randomly pick a user, user id 400, as the active test user. Before we evaluate wvec and ivec, we prepare each user's wvec by computing the average of all ivec of the movies the user has watched. The wvec of the active test user, user id 400, depicts in table 4, representing the overall average of 43 movies' ivec the user id 400 has watched.

Table 4: The average mood value wvec of user id 400 user.

| neutral | joy | sadness | hate | anger | disgust | surprise |
|---------|-----|---------|------|-------|---------|----------|
| 0.16352993 | 0.08873525 | 0.1270899 | 0.2033184 | 0.1193381 | 0.1588128 | 0.1391753 |

## 5. EVALUATION

We deployed the ml-latest-small dataset as the training dataset. We then randomly picked user id 400 as the active test user. We created the testing dataset from concatenating the other MovieLens datasets: ml-20m, ml-25m, and ml-latest-full. We extracted all data points that was belonging to the user id 400 and removed all the duplicated data points from the testing dataset and those found in the training dataset and named it as test400 dataset. The list of movies contains in the test400 dataset represents the list of movies the active user id 400 has yet watched. We compare the top-20 movie list generated by the Recommender algorithms against the active user unseen movie list in the testing dataset. We also get each top-5 list from each top-20 list by computing the closest similarity between the active user's wvec and each movie's ivec on the top-20 list and sorted in the descending order. In the top-5 list, films indicate a high probability the active user may accept one of the movies from the recommendations. However, the assumption we make on the active user choosing one of the unwatched films from the recommendations has a drawback. If a movie the active user likes to watch does not appear on the list, he would not choose the cinema but wait till he sees the popular film shows up on the recommendation list.

### 5.1. Top-20 Lists Generated by Recommenders

To generate movie recommendations for the active test user id 400, we chose a watched movie from the user's watched list, "Indiana Jones and the Last Crusade (1989)" as a basis. For each Recommender algorithm in the platform, we generated a top-20 movie recommendations list for the user id 400. We depicted in table 5, a collection of five top-20 movie recommendations lists made by each Recommender. Due to the limitation of space, table 4 will only show the top-20 list by movie id. For the corresponding movie titles, please refer to table 7 (A) and (B).

Table 5. Top-20 recommendations list generated by 5 recommenders for user id 400 user.

| No. | IBCF | UBCF | GAR | EAR | MAR |
|-----|------|------|-----|-----|-----|
| 1 | 1198 | 5952 | 761 | 7386 | 2879 |
| 2 | 2115 | 6016 | 90403 | 3283 | 112897 |
| 3 | 1196 | 4226 | 112897 | 3174 | 3283 |
| 4 | 1210 | 2329 | 32511 | 35807 | 5803 |
| 5 | 1036 | 2858 | 4367 | 6911 | 25946 |
| 6 | 1240 | 1089 | 160563 | 2243 | 2471 |
| 7 | 260 | 2762 | 115727 | 1496 | 1801 |

| 8  | 1270 | 68157  | 7925   | 7132   | 7302   |
| 9  | 2716 | 48394  | 1049   | 106920 | 122918 |
| 10 | 1200 | 110    | 3999   | 2017   | 2990   |
| 11 | 1214 | 2028   | 91485  | 144478 | 95510  |
| 12 | 1580 | 1682   | 147662 | 95441  | 5540   |
| 13 | 2571 | 115713 | 50003  | 43556  | 69278  |
| 14 | 589  | 1206   | 5244   | 96530  | 7248   |
| 15 | 1527 | 1704   | 112911 | 378    | 31923  |
| 16 | 1265 | 1      | 131714 | 2248   | 149406 |
| 17 | 1097 | 3147   | 3389   | 5088   | 134775 |
| 18 | 1136 | 1732   | 704    | 5667   | 112175 |
| 19 | 2028 | 27773  | 1606   | 2879   | 79695  |
| 20 | 1197 | 1228   | 4565   | 5803   | 5264   |

## 5.2. Top-5 Lists Extracted from Recommenders' Top-20 Lists

Using the wvec of the user id 400, we compute the pairwise similarity between the user id 400 and each recommended movie's ivec on the top-20 list. We sorted the pairwise distance metrics top-20 list calculated using wvec in the descending order to obtain the top-5 list for each of the Recommenders. Table 6 shows the computed top-5 list for each Recommender.

Table 6. Top-5 list computed via wvec of user id 400 and ivec of top-20 generated by the corresponding Recommenders.

| No. | IBCF | UBCF | GAR    | EAR    | MAR    |
|-----|------|------|--------|--------|--------|
| 1   | 2716 | 1732 | 90403  | 3174   | 122918 |
| 2   | 1527 | 2329 | 1606   | 43556  | 134775 |
| 3   | 2115 | 2858 | 5244   | 95441  | 112175 |
| 4   | 1240 | 6016 | 4367   | 5667   | 25946  |
| 5   | 1036 | 1206 | 147662 | 144478 | 79695  |

## 5.3. Limitations

Listed below are limitations that we observed in the study.

- Due to the lack of emotion labeled movie datasets, we cannot avow to the accuracy of the moods labeled movie dataset generated from the Tweets Affection Classifier. However, from our observation, TAC did a fair job of classifying films' emotional attributes from overviews.
- We adopted seven categories of emotion: the more classes we want to add to the collection, the harder to find adequate labeled data. For our purpose, seven seems to be stretching the limit.
- The top-N list that generates by each Recommender is unique. It presents a problem in finding an adequate evaluation metrics for benchmarking. The five top-20 recommendations lists that made by each respective Recommender are so different that we find ourselves comparing these lists as if comparing apples and oranges. They are all fruits but with very different tastes. For the time being, we rely on our intuition to judge how good these top-20 and top-5 lists are.

## 6. FUTURE WORK

We started the track to study the impact of affective features may have on Recommender Systems by examining how emotional attributes can interplay at the stage of the Recommender making top-N recommendations [1]. In this paper, we introduced a way to make Recommender emotionally aware. We focused on extracting affective features from textual movie metadata. We plan soon to perform an in-depth study in Multi-channel Emotion Aware Recommender by extracting emotion features from images such as movie posters as a component in building the Recommender. We also intrigued by the idea of using users' emotion profiles to enhance Group Recommenders in user grouping, group formation, group dynamics, and group decision making.

Table 7 (A). TopN recommendation list generated by five recommenders for user id 400 user.

| No. | Mid | Title |
|---|---|---|
| 0 | 1 | Toy Story (1995) |
| 1 | 110 | Braveheart (1995) |
| 2 | 260 | Star Wars: Episode IV - A New Hope (1977) |
| 3 | 378 | Speechless (1994) |
| 4 | 589 | Terminator 2: Judgment Day (1991) |
| 5 | 704 | Quest, The (1996) |
| 6 | 1036 | Die Hard (1988) |
| 7 | 1049 | Ghost and the Darkness, The (1996) |
| 8 | 1089 | Reservoir Dogs (1992) |
| 9 | 1097 | E.T. the Extra-Terrestrial (1982) |
| 10 | 1197 | Princess Bride, The (1987) |
| 11 | 1206 | Clockwork Orange, A (1971) |
| 12 | 1210 | Star Wars: Episode VI - Return of the Jedi (1983) |
| 13 | 1214 | Alien (1979) |
| 14 | 1240 | Terminator, The (1984) |
| 15 | 1265 | Groundhog Day (1993) |
| 16 | 1270 | Back to the Future (1985) |
| 17 | 1496 | Anna Karenina (1997) |
| 18 | 1527 | Fifth Element, The (1997) |
| 19 | 1580 | Men in Black (a.k.a. MIB) (1997) |
| 20 | 1682 | Truman Show, The (1998) |
| 21 | 1704 | Good Will Hunting (1997) |
| 22 | 1732 | Big Lebowski, The (1998) |
| 23 | 1801 | Man in the Iron Mask, The (1998) |
| 24 | 2017 | Babes in Toyland (1961) |
| 25 | 2243 | Broadcast News (1987) |
| 26 | 2248 | Say Anything... (1989) |
| 27 | 2471 | Crocodile Dundee II (1988) |
| 28 | 2571 | Matrix, The (1999) |
| 29 | 2716 | Ghostbusters (a.k.a. Ghost Busters) (1984) |
| 30 | 2762 | Sixth Sense, The (1999) |

Table 7 (B). Top-N recommendation list generated by five recommenders for user id 400 user.

| No. | Mid | Title |
|---|---|---|
| 31 | 2858 | American Beauty (1999) |
| 32 | 2990 | Licence to Kill (1989) |
| 33 | 3147 | Green Mile, The (1999) |

| 34 | 3389 | Let's Get Harry (1986) |
|----|------|------------------------|
| 35 | 3999 | Vertical Limit (2000) |
| 36 | 4367 | Lara Croft: Tomb Raider (2001) |
| 37 | 5088 | Going Places (Valseuses, Les) (1974) |
| 38 | 5244 | Shogun Assassin (1980) |
| 39 | 5540 | Clash of the Titans (1981) |
| 40 | 5667 | Tuck Everlasting (2002) |
| 41 | 6911 | Jolson Story, The (1946) |
| 42 | 7132 | Night at the Opera, A (1935) |
| 43 | 7248 | Suriyothai (a.k.a. Legend of Suriyothai, The) (2001) |
| 44 | 7302 | Thief of Bagdad, The (1924) |
| 45 | 7925 | Hidden Fortress, The (Kakushi-toride no san-akunin) (1958) |
| 46 | 25946 | Three Musketeers, The (1948) |
| 47 | 31923 | Three Musketeers, The (1973) |
| 48 | 43556 | Annapolis (2006) |
| 49 | 48394 | Pan's Labyrinth (Laberinto del fauno, El) (2006) |
| 50 | 50003 | DOA: Dead or Alive (2006) |
| 51 | 68157 | Inglourious Basterds (2009) |
| 52 | 69278 | Land of the Lost (2009) |
| 53 | 95441 | Ted (2012) |
| 54 | 96530 | Conception (2011) |
| 55 | 106920 | Her (2013) |
| 56 | 112175 | How to Train Your Dragon 2 (2014) |
| 57 | 112911 | Hercules (2014) |
| 58 | 115713 | Ex Machina (2015) |
| 59 | 115727 | Crippled Avengers (Can que) (Return of the 5 Deadly Venoms) (1981) |
| 60 | 122918 | Guardians of the Galaxy 2 (2017) |
| 61 | 131714 | Last Knights (2015) |
| 62 | 134775 | Dragon Blade (2015) |
| 63 | 147662 | Return of the One-Armed Swordsman (1969) |
| 64 | 149406 | Kung Fu Panda 3 (2016) |
| 65 | 160563 | The Legend of Tarzan (2016) |

## 7. CONCLUSION

Leverage on our prior work in affective computing [1] that making use of the Tweets Affective Classifier (TAC) to generate our needed movie emotion labeled dataset; we demonstrated in this paper a method to build an Emotion Aware Recommender (EAR) with intriguing results. We developed a Recommender platform using the following Recommender algorithms: Item-based Collaborative Filtering (IBCF), User-based Collaborative Filtering (UBCF), Content-based movie Genres Aware Recommender (GAR), Content-based Emotion Aware Recommender (EAR), and Content-based Multi-channel Emotion Aware Recommender (MAR). With each Recommender algorithm, we generate a top-20 recommendation list. We randomly selected user id 400 as an active user for testing. We compute the emotion profile, wvec, for the test user. Using the test user's wvec and the list of ivec from each of the top-20 list, we computed the top-5 for each top-20 list generated by the Recommenders. The top-N list made by each Recommender is unique, with few overlaps among the lists. We have a total of 100 movies in the combined top-20 lists. We found 35 duplicated films among the top-20 recommendation lists. The top-N list made by each Recommender met its design focus. For example, GAR correctly recommended movies that meet the active test user's genre taste. EAR, on the other hand, shows intrigue results. We believe that with further investigation, we could enhance EAR to make serendipity recommendations.

## AUTHORS

**John K. Leung** is a Ph.D. candidate in Computational and Data Sciences Department, Computational Sciences and Informatics at George Mason University in Fairfax, Virginia. He has over twenty years of working experience in information technology research and development capacity. Formerly, he worked in the T. J. Watson Research Center at IBM Corp. in Hawthorne, New York. John has spent more than a decade working in Greater China, leading technology incubation, transfer, and new business development.

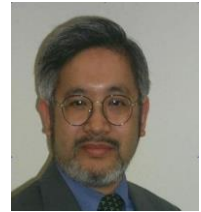

**Igor Griva** is an Associate Professor in the Department of Mathematical Sciences at George Mason University. His research focuses on the theory and methods of nonlinear optimization and their application to problems in science and engineering.

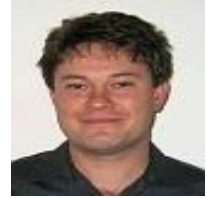

**William G. Kennedy**, PhD, Captain, USN (Ret.) is an Associate Professor in the Department of Computational and Data Sciences and is a Co-Director of the Center for Social Complexity at George Mason University in Fairfax, Virginia. He has over 10-years' experience in leading research projects in computational social science with characterizing the reaction of the population of a mega-city to a nuclear WMD event being his most recent project. His teaching, research, and publication activities are in modeling cognition and behavior from individuals to societies.

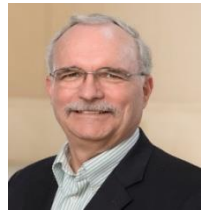